\documentclass{ieeeaccess}

\usepackage{cite}
\usepackage{amsmath,amssymb,amsfonts}
\usepackage{algorithmic}
\usepackage{graphicx}
\usepackage{textcomp}

\usepackage{url}
\usepackage{booktabs}
\usepackage{multirow}
\usepackage{threeparttable}
\usepackage{hyperref}

\usepackage{caption, subcaption}
\def\BibTeX{{\rm B\kern-.05em{\sc i\kern-.025em b}\kern-.08em
    T\kern-.1667em\lower.7ex\hbox{E}\kern-.125emX}}

\usepackage{xspace}

\newcommand{\proposedmodel}{DM$^2$S$^2$\xspace}

\newcommand{\CLS}{\texttt{[C]}}
\newcommand{\SEP}{\texttt{[S]}}

\begin{document}
\history{Received 31 August 2022, accepted 3 November 2022, date of publication 14 November 2022, date of current version 17 November 2022.
\doi{10.1109/ACCESS.2022.3221812}}

\title{
    DM$^2$S$^2$: Deep Multimodal Sequence Sets with Hierarchical Modality Attention
}

\author{
    \uppercase{Shunsuke Kitada}\authorrefmark{1},
    \uppercase{Yuki Iwazaki\authorrefmark{2}, Riku Togashi\authorrefmark{2}, and}
    \uppercase{Hitoshi Iyatomi}\authorrefmark{1}\IEEEmembership{Member, IEEE}
}
\address[1]{Department of Applied Informatics, Graduate School of Science and Engineering, Hosei University, Tokyo 184-8584, Japan}
\address[2]{CyberAgent, Inc., Tokyo 150-0042, Japan}
\tfootnote{
    This work was supported in part by JSPS KAKENHI under Grant 21J14143. The work of Shunsuke Kitada was supportedby CyberAgent Inc.
    \textit{\textcopyright 2022 IEEE. Personal use of this material is permitted. Permission from IEEE must be obtained for all other uses, in any current or future media, including reprinting/republishing this material for advertising or promotional purposes, creating new collective works, for resale or redistribution to servers or lists, or reuse of any copyrighted component of this work in other works.}
}

\markboth
{S. Kitada \headeretal: DM$^2$S$^2$: Deep Multimodal Sequence Sets with Hierarchical Modality Attention}
{S. Kitada \headeretal: DM$^2$S$^2$: Deep Multimodal Sequence Sets with Hierarchical Modality Attention}

\corresp{Corresponding author: Shunsuke Kitada (e-mail: shunsuke.kitada.8y@stu.hosei.ac.jp).}

\begin{abstract}
    There is increasing interest in the use of multimodal data in various web applications, such as digital advertising and e-commerce. 
Typical methods for extracting important information from multimodal data rely on a mid-fusion architecture that combines the feature representations from multiple encoders. 
However, as the number of modalities increases, several potential problems with the mid-fusion model structure arise, such as an increase in the dimensionality of the concatenated multimodal features and missing modalities. 
To address these problems, we propose a new concept that considers multimodal inputs as a set of sequences, namely, deep multimodal sequence sets (DM$^2$S$^2$). 
Our set-aware concept consists of three components that capture the relationships among multiple modalities: (a) a BERT-based encoder to handle the inter- and intra-order of elements in the sequences, (b) intra-modality residual attention (IntraMRA) to capture the importance of the elements in a modality, and (c) inter-modality residual attention (InterMRA) to enhance the importance of elements with modality-level granularity further. 
Our concept exhibits performance that is comparable to or better than the previous set-aware models. 
Furthermore, we demonstrate that the visualization of the learned InterMRA and IntraMRA weights can provide an interpretation of the prediction results.
\end{abstract}


\titlepgskip=-15pt

\maketitle

\section{Introduction}
    In industry (e.g., digital advertising~\cite{chen2016deep,zhang2018equal} and e-commerce~\cite{he2016ups,chen2019personalized}), huge amounts of multimodal data are often obtained, in which each instance is described as a set of information from multiple modalities.
Thus, handling multimodal information is a fundamental problem in most web applications.
Deep neural networks (DNNs) also experience technical challenges in handling multimodal data~\cite{baltruvsaitis2018multimodal}.

The mainstream approaches of multimodal models are based on a {\it mid-fusion} architecture, which encodes the data from each modality and then concatenates multiple features~\cite{arevalo2017gated,perez2019mfas}.
Previous studies have explored various mid-fusion models, such as the gated multimodal units (GMU)~\cite{arevalo2017gated} and neural architecture search (NAS)-based models~\cite{perez2019mfas}.
However, mid-fusion models often face several challenges.
First, as the number of modalities increases, the dimensionality of the concatenated features generally increases as well.
Moreover, mid-fusion models require the impractical assumption that each instance in a dataset must have the same number of modalities without missing information.

To address the above multimodal problems, Reiter~\textit{et al.}~\cite{reiter2020deep} proposed deep multimodal sets (DMMS) based on {\it Deep Sets}~\cite{zaheer2017deep}, which is a class of models that is defined according to sets.
The underlying concept of DMMS is that a function defined on sets should satisfy the invariance to the number and order of elements in a set (i.e., modalities).
However, as DMMS encodes/compresses each modality and then further compresses the features using a pooling operation, the architecture may omit the importance of signals in a single modality.

Although set-aware architectures such as DMMS~\cite{reiter2020deep} can deal with ``unordered’’ sets of modalities, the data in multimodal information takes various forms; certain types of data are inherently sequential (e.g., text tokens), whereas others are not (e.g., categorical data).
Recent BERT~\cite{devlin2019bert}-based multimodal methods often convert an image into textual tokens using optical character recognition (OCR), and thus, obtain a textual sequence~\cite{xu2020layoutlm,xu2021layoutlmv2,hong2022bros}.
In particular, such a preprocessing technique is effective in extracting rich semantic information that is difficult to capture by encoding raw modality data.
Therefore, these situations motivate us to handle {\it sets of sequences}.

As the interpretability and explainability of the model predictions are indispensable factors in industry, it is necessary to provide an interpretation of the importance of each modality.
The DMMS~\cite{reiter2020deep} enables an explainable prediction in terms of the importance of the modalities.
However, conventional mid-fusion models may ignore the higher-order interactions between elements across sequences, as they only focus on the fusion of encoded modality representations.
Furthermore, owing to the black-box nature of DNN-based models, such ``encode-and-concatenate'' models do not provide a breakdown of the importance of factors in a modality.
This is an insufficient diagnosis in a practical setting; for example, it is often desirable to know the element-level importance to investigate the problem in text tokenization.

In this study, we propose a new concept for handling multimodal data known as {\it deep multimodal sequence sets} (\proposedmodel).
Our concept consists of three components that capture the relationship between multiple modalities: (1) a BERT-based encoder, (2) intra-modality residual attention (IntraMRA), and (3) inter-modality residual attention (InterMRA).
For (1), the BERT-based encoder outputs a rich representation that handles the inter- and intra-orders of the elements in a modality.
In (2), IntraMRA captures the importance of the elements in a modality.
In (3), InterMRA further enhances the importance of the elements with modality-level granularity.
We employ OCR methods to extract semantic features from multimodal data (especially for the visual modality) that may not be captured by image recognition.
The model based on our concept empirically demonstrates that it is preferable to use semantic features that are derived from OCR tokens over features that are encoded directly from visual information.
Our two-step MRA can capture fine granularity representations that are not explicitly modeled by the DMMS~\cite{reiter2020deep}.

Empirical experiments on various real-world datasets reveal that our sequence set-aware concept achieved performance that is comparable to or better than previous multimodal and set-aware models.
In particular, the model based on our concept outperformed those deployed in the production environment on a large-scale production advertising (Ad) dataset.
Furthermore, the model enables the visualization of the contribution of elements in modalities to model predictions based on the importance that is learned by the MRA.

The contributions of this study are summarized as follows:
\begin{itemize}
    \item We propose a new model for capturing higher-order interactions, and the inter- and intra-relationships of each modality, considering that multimodal data often include set structured contents.
    \item Our set-aware model with a multimodal structure achieves equivalent or better prediction performance compared to previous multimodal and set-aware models.
    \item An evaluation on datasets, including a real-world production dataset, demonstrates that our \proposedmodel can be reasonably applied in a real environment. The visualization of the learned MRA weights can be used to explain predictions that are important in industrial fields.
\end{itemize}

\section{Related Work}
    \subsection{Multimodal Models}

In general, DNN-based multimodal models include multiple streams of networks for modalities~\cite{atrey2010multimodal}.
The models often have a component for fusing the features~\cite{wimmer2008low,arevalo2017gated,zhang2018equal} to make a prediction based on intermediate features from the streams.
Hence, by considering the fusion stage in the model architecture, multimodal architectures can be categorized into early fusion~\cite{wimmer2008low}, mid-fusion~\cite{atrey2010multimodal,fukui2016multimodal,arevalo2017gated, perez2019mfas}, and late fusion~\cite{zhang2018equal,wang2020makes}.

The early fusion approach integrates the features of different modalities as inputs and uses a unified feature for the downstream task~\cite{wimmer2008low}.
The mid-fusion approach involves the concatenation of features that are encoded from the raw data of each modality into a single feature~\cite{atrey2010multimodal,fukui2016multimodal,arevalo2017gated,perez2019mfas}.
Typical multimodal architectures rely on the mid-fusion approach to combine multimodal information.
However, mid-fusion may suffer from the increased dimensionality of concatenated features and the impractical assumption of complete modality information for each sample.
The late fusion approach combines the prediction scores of different predictors for multiple modalities~\cite{zhang2018equal,wang2020makes}.
The late fusion method often exhibits sub-optimality of the trained models owing to the dependency between modalities being ignored.

To solve this problem of mid-fusion architecture, which is the most common architecture in multimodal models, Reiter~\textit{et al.}~\cite{reiter2020deep} proposed the DMMS, which is a set-aware model based on Deep Sets~\cite{zaheer2017deep} that represents a collection of features as an unordered set rather than an increasing fixed-size vector.
The DMMS can deal with the set input by pooling an arbitrary number of modality features into a constant-size vector without dependence on the number of modalities.
Nevertheless, the DMMS may ignore the importance of signals in a single modality by compressing the modality information through the leading encoder.
Several mid-fusion models in the earlier stage of architecture introduced techniques to capture the higher-order interactions between modalities to construct each modality representation~\cite{gao2018question,liu2020learning}.
Owing to the set-aware architecture, the DMMS can handle an arbitrary number of modality features without any unnatural workarounds (e.g., padding for missing modalities).

This study presents an architecture for retaining the fine-grained signals in each modality while capturing the importance between the modalities.
Our proposed mid-fusion architecture operates similarly to models in the class of Deep Sets with modality-level granularity as in the DMMS while leveraging the hierarchical structure of multimodal data, namely, sets of sequences.
As recent BERT-based encoders output a sequence of embeddings rather than a single representation for an input sequence, we leverage it as a rich modality representation through our two-step attention mechanism.

\subsection{Modality-aware Attention Mechanism}

According to Bahdanau~\textit{et al.}~\cite{bahdanau2015neural}, various modality-aware attention mechanisms have been proposed~\cite{moon2018multimodal, zhou2019modality, liu2019roberta} to achieve effective learning in multimodal tasks.
Similar concepts have been explored for named entity recognition~\cite{moon2018multimodal}, audio-visual speech recognition~\cite{zhou2019modality}, and sentence summarization~\cite{li2018multi}.

This study employs an attention mechanism to capture the intra-modality signals and inter-modality importance based on the representation of sequence sets.
Whereas our proposed MRA is closely related to the modality attention proposed by Moon~\textit{et al.}~\cite{moon2018multimodal}, our framework clearly differs in that it considers the hierarchical structure, such as the inter- and intra-relationships of modalities. 
Furthermore, we introduce a residual connection to preserve the information of the source feature representation.

\begin{figure*}[t]
    \centering
    \includegraphics[width=\linewidth]{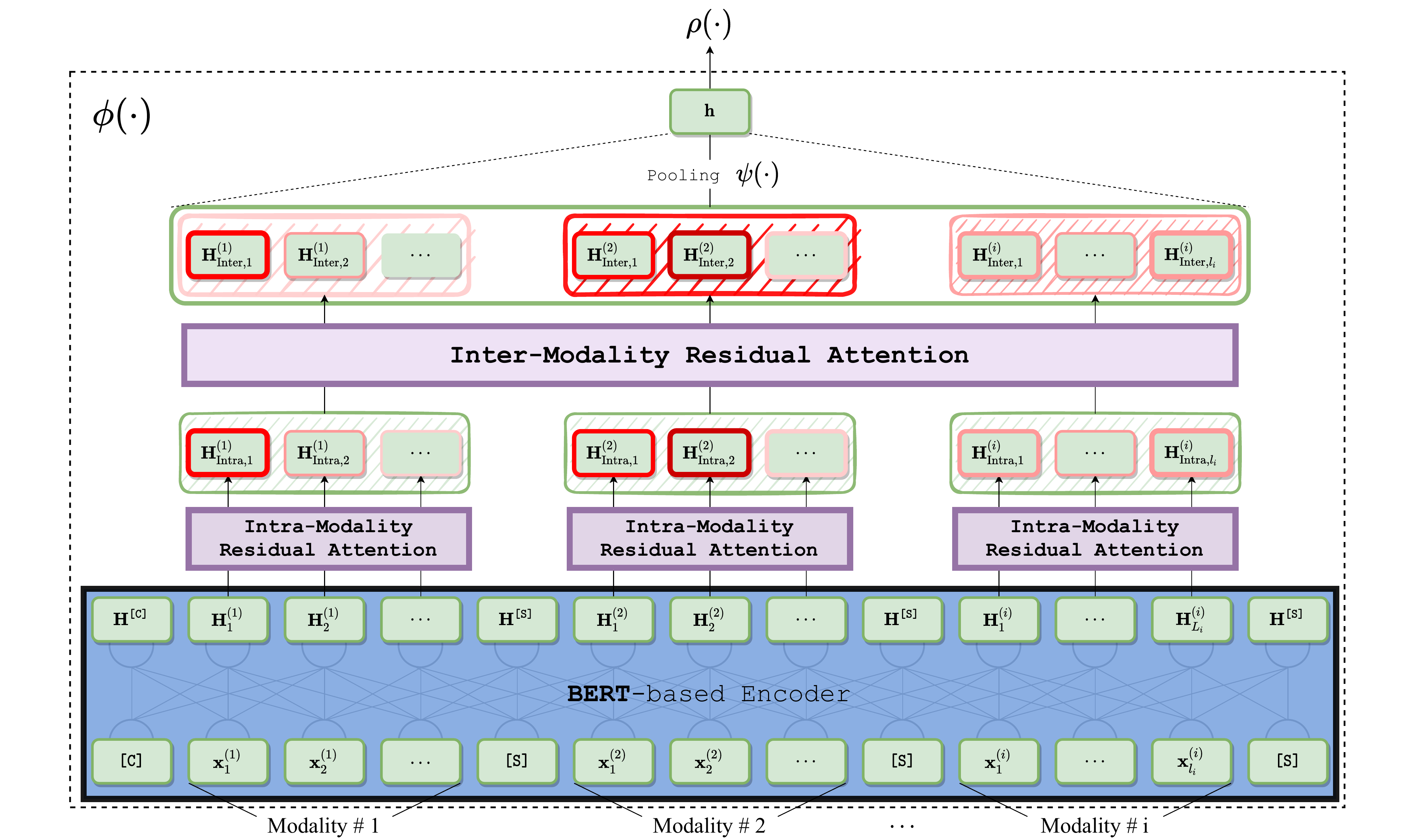}
    \caption{Model overview. The input is the sets of sequence from multiple modalities.
    The sequence is encoded into the hidden states by the BERT-based encoder.
    Our Intra- and InterMRA learn each relationships and their contribution (as indicated by the red field) of the modalities.
    The output is a hidden vector that is obtained through the encoder and the mechanisms.
    }
    \label{fig:overview}
\end{figure*}

\section{Methodology}
    First, we briefly introduce Deep Sets.
Our \proposedmodel~is based on Deep Sets, which is a class of models defined according to sets.
Subsequently, we describe the proposed \proposedmodel and its key component, namely MRA.

\subsection{Deep Sets}

For each modality $m \in \mathcal{M}$, we extract a sequence $\mathbf{x}^{(m)} \in \mathcal{V}^{l_m}$ of length $l_m$ from the tokens that are contained in vocabulary set $\mathcal{V}$.
Each token is expected to be extracted appropriately depending on the modality type; for example, a word or character for the textual modality and an OCR-detected word for the visual modality.
Moreover, tokens can be obtained from images, audio sources, and videos; that is, a visual patch such as ViT~\cite{dosovitskiy2020image} from an image, transcription from audio, and the results of the same procedure for images from each frame of video.

We wish to handle the set of sequences $X=\{\mathbf{x}^{(m)} \mid m \in \mathcal{M}\}$ as the input of a model; the loss functions should also be computed on such sets.
To this end, we design our proposed architecture based on Deep Sets~\cite{zaheer2017deep} as in the DMMS~\cite{reiter2020deep}.
A model in the class of Deep Sets can be expressed as follows:
\begin{equation}
    f(X) = \rho(\phi(X)),
\end{equation}
where $\phi$ is the neural network that encodes the input set $X$, and $\rho$ is the prediction network for a downstream task.

As described above, we use the sets of sequences that are extracted from each modality as inputs for the model.
Therefore, in contrast to the conventional models~\cite{zaheer2017deep,reiter2020deep}, we aim to handle the order between the tokens $\{\mathbf{x}_i^{(m)} \mid 1\leq i \leq l_m\}$ of each modality $m$, but ignore that between the modalities $m \in \mathcal{M}$.

\subsection{Deep Multimodal Sequence Sets (\proposedmodel)}

Fig.~\ref{fig:overview} presents an overview of our concept.
The encoder $\phi$ comprises three components: (1) a BERT-based encoder; (2) IntraMRA, and
(3) InterMRA.
The input is the set of sequences from multiple modalities.
Each set contains token representations, as described previously.
Even if this unimodalized (tokenized) information is input, we consider this to be a multimodal setting.

First, to handle the sets of sequences appropriately, a BERT-based encoder is used as a building block of $\phi$.
We construct a unified sequence by concatenating the sequences of the modalities:
\begin{equation}
    \mathbf{x} = \biggl\{
        \begin{aligned}
            &\CLS, \mathbf{x}_1^{(1)}, \cdots, \mathbf{x}_{l_1}^{(1)}, \SEP, \cdots, \\
            &\hspace{.5em}   
            \SEP, \mathbf{x}_1^{(M)}, \cdots, \mathbf{x}_{l_M}^{(M)}, \SEP
        \end{aligned}
    \biggr\} \in \mathbb{\mathcal{V}}^{L+M},
\end{equation}
where $\mathbf{x}^{(m)}_i \in \mathcal{V}$ indicates the $i \in [l_m]$-th token in the sequence of the modality $m \in \mathcal{M}$.  
The first dimensionality of $\mathbf{x}$ originates from the total number $L$ of modality tokens in $\mathbf{x}$ and the number of the BERT special tokens (i.e., one $\CLS$ token and $M \SEP$ tokens).
Hence, we consider $\CLS$ and $\SEP$ as elements in the vocabulary set (i.e., $\CLS, \SEP \in \mathcal{V}$).
We obtain the representation $\mathbf{H}^{(m)}_i \in \mathbb{R}^d$ as a $d$-dimensional hidden state of the BERT encoder.
Thus, the token sequence $\mathbf{x}^{(m)}$ of modality $m$ is transformed into the BERT embedded sequence $\mathbf{H}^{(m)} = (\mathbf{H}^{(m)}_1, \cdots, \mathbf{H}^{(m)}_{l_m}) \in \mathbb{R}^{l_m \times d}$.

We assume that each sample has the same number of modalities (i.e., $M$).
When a modality is missing for a sample, the corresponding entry ${\bf x}^{(m)}$ can be filed with an empty sequence.
Hence, by ensuring a fixed order of modalities in $\mathbf{x}$, the BERT encoder can be viewed as permutation-invariant in terms of the order of the modalities, and permutation-sensitive in terms of that of the tokens in a single modality.

As the BERT encoder can handle the order of tokens appropriately, we further enhance both the intra- and inter-modality relationships.
Intuitively, the importance of each token from multiple modalities can be decomposed into inter- and intra-modality factors; a token may be important because it is an essential modality for a prediction, and/or it is representative of its modality. 
We introduce a two-step MRA into the representation of $\mathbf{H}^{(m)}$ to model this hierarchical structure of token importance.

We apply $\mathrm{IntraMRA} \colon \mathbb{R}^{l_m \times d} \mapsto \mathbb{R}^{l_m \times d}$ to quantify the token importance in each modality $m \in \mathcal{M}$ (described in detail in Section~\ref{sec:method_intra_mra}).
\begin{equation}
    \mathbf{H}^{(m)}_{\mathrm{Intra}} = \mathrm{IntraMRA}(\mathbf{H}^{(m)}).
\end{equation}

We denote the concatenated matrix of the IntraMRA outputs for each modality as $\mathbf{H}_{\mathrm{Intra}} = (\mathbf{H}_{\mathrm{Intra}}^{(1)}, \cdots, \mathbf{H}_{\mathrm{Intra}}^{(M)}) \in \mathbb{R}^{L \times d}$, where $L = \sum_{m \in \mathcal{M}} l_m$.
Subsequently, we apply $\mathrm{InterMRA} \colon \mathbb{R}^{L \times d} \mapsto \mathbb{R}^{L \times d}$ to capture the importance of each modality (described in detail in Section~\ref{sec:method_inter_mra}):
\begin{equation}
    \mathbf{H}_{\mathrm{Inter}} = \mathrm{InterMRA}(\mathbf{H}_{\mathrm{Intra}}).
\end{equation}

The representation $\mathbf{H}_{\mathrm{Inter}}$ is aggregated as the final hidden vector $\mathbf{h} \in \mathbb{R}^d$ through a pooling function $\psi \colon \mathbb{R}^{L \times d} \mapsto \mathbb{R}^{d}$.
In this study, we use the following mean pooling as $\psi$:
\begin{equation}
    \mathbf{h} = \psi(\mathbf{H}_{\mathrm{Inter}}) = \frac{1}{L}\sum_{i=1}^{L} \mathbf{H}_{\mathrm{Inter}, i},
\end{equation}
where $\mathbf{H}_{\mathrm{Inter}, i}$ is the $i$-th token representation of $\mathbf{H}_{\mathrm{Inter}}$.

The projection that is defined by the above procedure can be considered as a permutation-agnostic encoder $\phi$.
This is because the Intra/InterMRAs and $\psi$ are evidently permutation invariant in terms of the order of the modalities.
Based on the representation of a sequence set $\phi(X)=\mathbf{h}$, we obtain the prediction $\hat{\mathbf{y}}=f(X)$ through a multi-layer perceptron (MLP), $\rho$:
\begin{equation}
    \hat{\mathbf{y}}= \rho(\mathbf{h}).
\end{equation}

\subsection{Modality Residual Attention (MRA)}

\subsubsection{Attention mechanism}

Our concept of MRA originates from the attention mechanism~\cite{bahdanau2015neural}.
The attention mechanism estimates the contribution of the input to the prediction.
Using these properties, we propose an MRA that can learn the intra- and inter-relationships for multiple modalities.
A possible extension is available to replace our Intra- and InterMRA mechanisms by adding a Transformer module~\cite{vaswani2017attention} to the BERT encoder; however, we will leave this as future work.

We compute the attention score from the alignment function $A \colon \mathbb{R}^{l} \mapsto [0, 1]^{l}$ by passing the hidden state ${\bf H} \in \mathbb{R}^{l \times d}$ to the attention mechanism, which consists of a similarity function $S \colon \mathbb{R}^{l} \mapsto \mathbb{R}^{l}$ followed by the softmax function:
\begin{equation}
    A_t(\mathbf{H}; \mathbf{W}, \mathbf{q}) = \mathrm{softmax}(S(\mathbf{H}_t; 
    \mathbf{W}, \mathbf{q})),
\end{equation}
where $\mathbf{W} \in \mathbb{R}^{d \times d}$ is a trainable weight matrix, and $\mathbf{q} \in \mathbb{R}^{d}$ is a trainable self-attention vector~\cite{yang2016hierarchical}.
The similarity function may be the additive attention~\cite{bahdanau2015neural} or the scaled dot-product attention~\cite{vaswani2017attention}.
In this study, we considered the additive attention, as follows:
\begin{equation}
    S(\mathbf{H}_t; \mathbf{W}, \mathbf{q}) = \mathbf{q}^\top\mathrm{tanh}(\mathbf{W}\mathbf{H}_t).
\end{equation}
We demonstrate that additive attention is preferable to scaled dot-product attention in Section~\ref{sec:ablation_similarity_function}.

\subsubsection{IntraMRA}\label{sec:method_intra_mra}

IntraMRA learns the token-level attention score for the modality $m$.
Although it is possible to learn the global attention (i.e., the attention of the entire token sequence) for the token sequence that is input into the encoder, the attention score for each token will be relatively small.
IntraMRA helps to learn the modality inter-relationships by calculating the attention for the token sequence in the modality.
We ensure that the value of the feature representation does not become excessively small by employing a residual connection~\cite{he2016deep} that preserves the information of the source feature representation.

For the hidden state $\mathbf{H}^{(m)}_t$ that is obtained by the encoder, the IntraMRA score $a_{\mathrm{Intra}, t}^{(m)}$ is calculated as follows:
\begin{equation}
    a_{\mathrm{Intra}, t}^{(m)} = A_t(\mathbf{H}^{(m)}; \mathbf{W}^{(m)}, \mathbf{q}^{(m)}),
\end{equation}
where $\mathbf{W}^{(m)}\in \mathbb{R}^{d \times d}$ and $\mathbf{q}^{(m)} \in \mathbb{R}^{d}$ are trainable parameters for modality $m$.
Using the attention score, we obtain the intra-attended representation $\mathbf{H}_{\mathrm{Intra}, t}^{(m)}$ with the residual function $F$:
\begin{equation}
    \mathbf{H}_{\mathrm{Intra}, t}^{(m)} = F(\mathbf{H}^{(m)}_t) + \mathbf{H}^{(m)}_t,
\end{equation}
where $F(\mathbf{H}^{(m)}_t) = a^{(m)}_{\mathrm{Intra}, t} \mathbf{H}^{(m)}_t$.
IntraMRA outputs the concatenated matrix for each element:
\begin{equation}
    \mathbf{H}_{\mathrm{Intra}}^{(m)} = \mathrm{IntraMRA}(\mathbf{H}^{(m)}) = (\mathbf{H}_{\mathrm{Intra}, 1}^{(m)}, \cdots, \mathbf{H}_{\mathrm{Intra}, l_m}^{(m)}).
\end{equation}

\subsubsection{InterMRA}\label{sec:method_inter_mra}

InterMRA learns the modality-level (e.g., visual-, textual-, and categorical-level) attention scores.
The mechanism captures the modalities that contribute to the prediction, which is important for model training.
We employ a residual connection for InterMRA with the expectation that the connection will provide the same effect as in IntraMRA.

We first compute the summation of each modality representation for $\mathbf{H}' \in \mathbb{R}^{M \times d}$ that contains the hidden representation for each modality $\mathbf{H}_{\mathrm{Intra}}$:
\begin{equation}
    \mathbf{H}' = \left(\sum_{t=1}^{l_1} \mathbf{H}_{\mathrm{Intra}, t}^{(1)}, \cdots, \sum_{t=1}^{l_m} \mathbf{H}_{\mathrm{Intra}, t}^{(m)}, \cdots, \sum_{t=1}^{l_M} \mathbf{H}_{\mathrm{Intra}, t}^{(M)}\right).
\end{equation}
Thereafter, we calculate the InterMRA score $a_{\mathrm{Inter}}^{(m)}$ for the modality $m$:
\begin{equation}
    a_{\mathrm{Inter}}^{(m)} = A_m(\mathbf{H}'; \mathbf{W}', \mathbf{q}'),
\end{equation}
where $\mathbf{W}' \in \mathbb{R}^{d \times d}$ and $\mathbf{q}' \in \mathbb{R}^{d}$ are trainable parameters for InterMRA.
We obtain the following inter-attended representation: $\mathbf{H}^{(m)}_{\mathrm{Inter}}$ from the InterMRA score $a_{\mathrm{Inter}}^{(m)}$ for modality $m$ with residual function $F'$:
\begin{equation}
    \mathbf{H}_{\mathrm{Inter}}^{(m)} = F'(\mathbf{H}_{\mathrm{Intra}}^{(m)}) + \mathbf{H}_{\mathrm{Intra}}^{(m)},
\end{equation}
where $F'(\mathbf{H}_{\mathrm{Intra}}^{(m)}) = a_{\mathrm{Inter}}^{(m)} \mathbf{H}_{\mathrm{Intra}}^{(m)}$
Finally, InterMRA outputs the resulting concatenated matrix for each modality, as follows:
\begin{equation}
    \mathbf{H}_{\mathrm{Inter}} = \mathrm{InterMRA}(\mathbf{H}_{\mathrm{Intra}}) = (\mathbf{H}_{\mathrm{Inter}}^{(1)}, \cdots, \mathbf{H}_{\mathrm{Inter}}^{(M)}).
\end{equation}

\section{Experiments}
    \subsection{Multimodal Datasets for Evaluation}

We used three multimodal datasets, namely, MM-IMDB~\cite{arevalo2017gated}, Ads-Parallelity~\cite{zhang2018equal}, and Production Ad-LP datasets for the empirical experiments.
The MM-IMDB~\cite{arevalo2017gated} dataset contains 25,925 movies with multiple labels (genres).
We used the original split provided in the dataset and reported the F1 scores (micro, macro, and samples) of the test set.
The Ads-Parallelity~\cite{zhang2018equal} dataset contains 670 images and slogans from persuasive advertisements to understand the implicit relationship (parallel and non-parallel) between these two modalities.
A binary classification task is used to predict whether the text and image in the same ad convey the same message.
We reported the overall and per-class average accuracy and ROC-AUC scores of five-fold cross-validation~\cite{zhang2018equal, reiter2020deep}.

The Ad-LP dataset contains 257,235 search engine ads and landing pages (LPs) with search keywords, ad titles, descriptions, and LP text.
The dataset was collected from CyberAgent Inc.\footnote{\url{https://www.cyberagent.co.jp/en/corporate/}} from August 1, 2020, to November 30, 2020.
The task is to regress the conversion rate from the LP after clicking on the ad; the conversion rate expresses the number of clicks/total number of impressions of the ad.
We divided the dataset into training, validation, and testing sets.
The resulting splits comprised 179,922, 45,138, and 32,175 Ad-LP pairs, respectively.

\subsection{Input Features}\label{sec:input_features}

We transformed the following multimodal information (i.e., visual, textual, and categorical data) into textual tokens and fed these into our proposed model.
We used the Google Cloud Vision API\footnote{\url{https://cloud.google.com/vision}} for the visual features to obtain the following four pieces of information as tokens\footnote{Both the MM-IMDB and Ads-Parallelity datasets, which include the cloud-based features, are available for download from the Zenodo platform: \url{https://doi.org/10.5281/zenodo.7050924}}: (1) text from the OCR, (2) category labels from the label detection, (3) object tags from the object detection, and (4) the number of faces from the facial detection.
We input the labels and object detection results as a sequence in order of confidence, as obtained from the API.
We describe the visual, textual, and categorical features of each dataset below.

\paragraph{\textbf{MM-IMDB}} We used the title and plot of movies as the textual features, and
the aforementioned API results are based on poster images as visual features.

\paragraph{\textbf{Ads-Parallelity}} We used the same API-based visual features as in MM-IMDB.
Furthermore, we used textual and categorical features consisting of textual inputs of transcriptions and messages, and categorical inputs of natural and text-concrete images.

\paragraph{\textbf{Production Ad-LP}} We used the following features: (1) as visual features, we used the result of applying the API to the screenshot for the first view of the LPs; (2) as textual features, we used the search words, titles, and descriptions of the ads, and the URL path for the LPs; and (3) as categorical features, we used the match type to search keywords as tokens.
Refer to Appendix~\ref{sec:details_of_production_ad_lp} for details on the production dataset.

\subsection{Implementation Details}

We implemented our proposed concept using AllenNLP version 2.5.0~\cite{gardner2018allennlp} with Hugging Face transformer version 4.5.1~\cite{wolf2020transformers}.
We evaluated the test set only once in all experiments.
The experiments were conducted on an Ubuntu 20.04 PC with an NVIDIA RTX A6000 GPU.

Following the DMMS of Reiter~\textit{et al.}~\cite{reiter2020deep}, we used a pre-trained RoBERTa large encoder~\cite{liu2019roberta}\footnote{\url{https://huggingface.co/roberta-large}} in MM-IMDB and Ads-Parallelity as the BERT-based encoder.
The encoder outputs the hidden state with dimension $d = 1024$.
We froze the parameters of the encoder and used the output of the hidden representation.
We used a pre-trained Japanese BERT\footnote{\url{https://huggingface.co/cl-tohoku/bert-base-japanese-whole-word-masking} } as the encoder in the Production Ad-LP dataset.
This encoder outputs a hidden state with dimension $d = 768$.

We followed the settings of Reiter~\textit{et al.}~\cite{reiter2020deep} for the optimizer.
Specifically, we used Adam~\cite{kingma2014adam} with a decoupled weight decay~\cite{loshchilov2018decoupled}.
The learning rate was warmed up linearly from 0 to 0.001 during the first five epochs and then decayed with a cosine annealing schedule over the remaining epochs.
For the MM-IMDB and Ads-Parallelity datasets, which are designed for single- and multi-label classification tasks, we used the sigmoid cross-entropy with class weights~\cite{king2001logistic} to train the model on $N$ training samples:
\begin{equation}
    \mathcal{L}_{\mathrm{WSCE}} = - \frac{1}{N} \sum_{j=1}^{N}\sum_{k=1}^{K} w_k \mathbf{Y}_{jk} \log (\hat{\mathbf{Y}}_{jk}),
\end{equation}
where $K$ is the number of classes and $w_k = 1/N_k$ is the class weight for class $k$ that is calculated with the class frequency $N_k$.
In this case, $\mathbf{Y} \in \{0, 1\}^{N \times K}$ is the target label matrix in which the $(j,k)$-entry indicates whether the training sample $j$ is in class $k$, and $\hat{\mathbf{Y}}\in [0, 1]^{N \times K}$ is the predicted label matrix.
For the Production Ad-LP dataset, which is a regression task, we used the root mean squared error (RMSE) to train the model, as follows:
\begin{equation}
    \mathcal{L}_{\mathrm{RMSE}} = \sqrt{\frac{\sum_{i=1}^{N} (y_i - \hat{y}_i)^2}{N}},
\end{equation}
where $y_i$ is the $i$-th ground truth and $\hat{y}_i$ is the $i$-th predicted value.
Refer to Appendix~\ref{sec:appendix_implementation_details} for the implementation details.

\section{Results and Discussion}
\begin{table*}[t]
\centering
\caption{
    Comparison of GMU, Bilinear-Gated, MFAS, MMBT, DMMS, and \proposedmodel on the MM-IMDB dataset. Compared to state-of-the-art models that handle multiple input formats (i.e., images and text encoded by their respective encoders), the proposed method that handles a single input format (i.e., multiple modalities as text) achieves comparable or better prediction performance.
}
\label{tab:sota_mm_imdb}
\begin{tabular}{@{}llrrr@{}}
\toprule
\multicolumn{1}{c}{Models}               & \multicolumn{1}{c}{OCR method}     & \multicolumn{1}{c}{F1-micro} & \multicolumn{1}{c}{F1-macro} & \multicolumn{1}{c}{F1-samples} \\ \cmidrule(r){1-1} \cmidrule(lr){2-2} \cmidrule(lr){3-3} \cmidrule(lr){4-4} \cmidrule(l){5-5}
GMU~\cite{arevalo2017gated}              & -                                  & 63.00                        & 54.10                        & 63.00                          \\
Bilinear-Gated~\cite{kiela2018efficient} & -                                  & 62.30                        & -                            & -                              \\
MFAS~\cite{perez2019mfas}                & -                                  & -                            & 55.68                        & -                              \\
MMBT~\cite{kiela2019supervised}          & -                                  & 66.40                        & 61.10                        & -                              \\
DMMS~\cite{reiter2020deep}               & Rosetta~\cite{borisyuk2018rosetta} & 67.73                        & 61.33                        & 67.63                          \\
                                         & Cloud Vision API                   & 67.97                        & 61.47                        & 67.86                          \\ \cmidrule(r){1-1} \cmidrule(lr){2-2} \cmidrule(lr){3-3} \cmidrule(lr){4-4} \cmidrule(l){5-5}
RoBERTa encoder only                     & Cloud Vision API                   & 64.37                        & 55.27                        & 64.24                          \\ \cmidrule(r){1-1} \cmidrule(lr){2-2} \cmidrule(lr){3-3} \cmidrule(lr){4-4} \cmidrule(l){5-5}
\proposedmodel (\textbf{proposed})                & Tesseract~\cite{kay2007tesseract}  & 59.48                        & 54.60                        & 58.92                          \\
                                         & EasyOCR                            & 66.86                        & 61.33                        & 66.59                          \\
                                         & Cloud Vision API                   & \textbf{69.64}               & \textbf{63.89}               & \textbf{69.33}                 \\ \bottomrule
\end{tabular}
\end{table*}
\begin{table*}[t]
\centering
\caption{
    Comparison of Combined Classifiers, DMMS, and \proposedmodel on the Ads-Parallelity dataset. The proposed model achieved performance comparable to or better than classifiers combining unimodal weak learners and DMMS, the state-of-the-art multimodal model. We observed that even though our framework uses only a single input format, it performs significantly better than DMMS, which uses multiple input formats.
}
\label{tab:sota_ads_parallelity}
\begin{tabular}{@{}llrrrrr@{}}
\toprule
\multicolumn{1}{c}{\multirow{2}{*}{Models}} & \multicolumn{1}{c}{\multirow{2}{*}{OCR method}} & \multicolumn{2}{c}{Overall}                            & \multicolumn{1}{c}{\multirow{2}{*}{\begin{tabular}[c]{@{}c@{}}Non-\\ Parallel\end{tabular}}} & \multicolumn{2}{c}{Parallel}                              \\ \cmidrule(lr){3-4} \cmidrule(l){6-7} 
\multicolumn{1}{c}{}                        & \multicolumn{1}{c}{}                            & \multicolumn{1}{c}{Accuracy} & \multicolumn{1}{c}{AUC} & \multicolumn{1}{c}{}                                                                         & \multicolumn{1}{c}{Equiv} & \multicolumn{1}{c}{Non-Equiv} \\ \cmidrule(r){1-1} \cmidrule(lr){2-2} \cmidrule(lr){3-3} \cmidrule(lr){4-4} \cmidrule(lr){5-5} \cmidrule(lr){6-6} \cmidrule(l){7-7}
Combined Classifiers~\cite{zhang2018equal}  & Cloud Vision API                                & 65.50                        & -                       & 63.30                                                                                        & 70.20                     & 65.50                         \\
DMMS~\cite{reiter2020deep}                  & Rosetta~\cite{borisyuk2018rosetta}              & 76.71                        & 77.79                   & 75.15                                                                                        & 88.97                     & 67.45                         \\
                                            & Cloud Vision API                                & 76.83                        & 77.98                   & 75.24                                                                                        & 89.03                     & 66.22                         \\ \cmidrule(r){1-1} \cmidrule(lr){2-2} \cmidrule(lr){3-3} \cmidrule(lr){4-4} \cmidrule(lr){5-5} \cmidrule(lr){6-6} \cmidrule(l){7-7}
RoBERTa encoder only                        & Cloud Vision API                                & 64.87                        & 65.83                   & 62.19                                                                                        & 70.11                     & 62.31                         \\ \cmidrule(r){1-1} \cmidrule(lr){2-2} \cmidrule(lr){3-3} \cmidrule(lr){4-4} \cmidrule(lr){5-5} \cmidrule(lr){6-6} \cmidrule(l){7-7}
\proposedmodel (\textbf{proposed})                   & Tesseract~\cite{kay2007tesseract}               & 66.76                        & 67.73                   & 63.51                                                                                        & 70.82                     & 65.95                         \\
                                            & EasyOCR                                         & 76.04                        & 76.08                   & 73.63                                                                                        & 88.81                     & 65.67                         \\
                                            & Cloud Vision API                                & \textbf{78.15}               & \textbf{79.22}          & \textbf{75.45}                                                                               & \textbf{91.52}            & \textbf{67.48}                \\ \bottomrule
\end{tabular}
\end{table*}

\subsection{Comparison with Baselines}

For MM-IMDB and Ads-Parallelity, we compared our model, \proposedmodel, and the state-of-the-art baselines based on NAS, Transformer, and set-aware models with a BERT-based encoder.
Tables~\ref{tab:sota_mm_imdb} and \ref{tab:sota_ads_parallelity} present the evaluation results for MM-IMDB and Ads-Parallelity, respectively.
We refer to the scores provided in each of the papers as a comparison and report them along with the \proposedmodel scores.

For MM-IMDB, we compared our \proposedmodel with five state-of-the-art models: GMU~\cite{arevalo2017gated}, Bilinear-Gated~\cite{kiela2018efficient}, multimodal fusion architecture search (MFAS)~\cite{perez2019mfas}, multimodal bitransformer (MMBT)~\cite{kiela2019supervised}, and DMMS~\cite{reiter2020deep}.
In addition to the baseline above baselines, we compared our model with a pure BERT-based encoder (indicated in the table as RoBERTa encoder only) that uses token type IDs (also known as segment IDs) to differentiate tokens that belong to different modalities. 
This is a common approach in the multimodal literature~\cite{zhou2020unified}.

As indicated in Table~\ref{tab:sota_mm_imdb}, \proposedmodel consistently outperformed the baselines on all performance measures.
The comparison between \proposedmodel and the NAS-based MFAS reveals that our \proposedmodel exhibits the advantage of token- and modality-level fusion based on the proposed two-step MRA, in contrast to MFAS, which searches for the best mid-fusion architecture based on the NAS algorithm.
As the DMMS is closely related to our \proposedmodel in terms of the set-aware fusion technique and BERT-based encoder, the performance gain of our \proposedmodel compared to the DMMS was remarkable.
However, it should be noted that the DMMS uses a proprietary OCR method, and, thus, a fair comparison between \proposedmodel and the DMMS is difficult owing to the difference in the OCR performances. 
Therefore, we reproduced the DMMS as effectively as possible and trained the model on a dataset built with tesseract~\cite{kay2007tesseract}, EasyOCR~\cite{easyocr} and the Cloud Vision API.

According to Table~\ref{tab:sota_mm_imdb}, \proposedmodel with the Cloud Vision API exhibited better results than the other OCR methods.
The performance of \proposedmodel with EasyOCR, which is a neural-based model based on CRAFT~\cite{baek2019character} and CRNN~\cite{shi2016end}, was comparable to that of the DMMS.
Although we conducted experiments using the same OCR methods (Cloud Vision API), our \proposedmodel outperformed the DMMS.
We also obtained good results in all performance measures for the RoBERTa encoder only model with token-type IDs.
Based on these results, we conclude that our MRA makes a more important contribution to learning multimodal data than the pure BERT-based model.

For Ads-Parallelity as indicated in Table~\ref{tab:sota_ads_parallelity}, we compared \proposedmodel with Combined Classifiers~\cite{zhang2018equal} and the DMMS~\cite{reiter2020deep}.
Furthermore, we compared the RoBERTa encoder only model, as described above.
In terms of all measures, the DMMS and \proposedmodel (with the Cloud Vision API) substantially outperformed Combined Classifiers, which uses features that are carefully designed for this dataset; moreover, a visual feature in Combined Classifiers is based on Cloud Vision API.
This demonstrates the effectiveness of the BERT-based encoder and set-aware architectures.
Despite the underperformance of Combined Classifiers with visual features from the Cloud Vision API compared to \proposedmodel and the DMMS, our \proposedmodel outperformed the DMMS when using the Cloud Vision API, while achieving comparable performance, even with EasyOCR.
The RoBERTa encoder only model, which does not include the MRA variants of \proposedmodel, could not outperform our \proposedmodel.

In summary, our \proposedmodel significantly outperformed the conventional methods on public datasets.
Compared with the baseline models that encode images directly, our model, which tokenizes even images, consistently exhibited superior results.
We believe that our model can capture more semantic features by employing the OCR method to extract the features explicitly, instead of implicitly, from images.
Our model is expected to be effective for modalities other than images, such as audio and video, in which text tokens can be obtained via transcription.
Compared to the DMMS, which is based on a set-aware architecture and a BERT-based encoder, our \proposedmodel outperformed the baselines on all evaluation measures.
Considering that the DMMS uses proprietary visual features, we also examined the effect of the OCR methods in \proposedmodel as far as possible.

\subsection{In-Depth Analysis}

\subsubsection{Ablation Study on Intra and InterMRA}

\begin{table}[t]
    \centering
    \caption{Performance comparison of MRA variants.}
    \begin{tabular}{@{}lrrrr@{}}
        \toprule
        \multirow{2}{*}{MRA} & \multicolumn{2}{c}{MM-IMDB}  &   \multicolumn{2}{c}{Ads-Parallelity} \\
        \cmidrule(r){2-3} \cmidrule(l){4-5} &   \multicolumn{1}{c}{\small F1-micro} &   \multicolumn{1}{c}{\small F1-macro} &   \multicolumn{1}{c}{\small Acc.} &   \multicolumn{1}{c}{\small AUC}  \\
        \cmidrule(r){1-1} \cmidrule(lr){2-2} \cmidrule(lr){3-3} \cmidrule(lr){4-4} \cmidrule(l){5-5}
        RoBERTa encoder only &   63.32   &   57.57   &   71.70   &   72.80   \\
        + IntraMRA          &   68.32   &   62.57   &   76.70   &   77.80   \\
        + InterMRA          &   67.27   &   60.87   &   72.16   &   72.98   \\
        + Intra \& InterMRA    &   \textbf{69.64}  &   \textbf{63.89}  &   \textbf{78.15}  &   \textbf{79.22}  \\
        \bottomrule
    \end{tabular}
    \label{tab:lma_hma}
\end{table}

Table~\ref{tab:lma_hma} displays the comparison of the prediction performance of MRA when using both Inter- and IntraMRA.
Although InterMRA and IntraMRA are beneficial, even when they are used independently, \proposedmodel achieved the best performance by leveraging them simultaneously.
In contrast, the performance gain of IntraMRA was slightly larger than that of InterMRA.
This is possibly because inter-modality importance is essential for capturing several important tokens in a sequence when the number of tokens is large.
However, InterMRA can quantify the importance of modalities rather than that of tokens, and is thus complementary to IntraMRA.

We further examined the effectiveness of our concept on a real-world dataset, namely Production Ad-LP.
As a baseline model for the Production Ad-LP dataset, we considered a gradient boosting decision tree~\cite{friedman2001greedy} model that provides daily predictions of conversion rates from ads and LPs in the production environment at CyberAgent Inc. 
The model was implemented by using LightGBM~\cite{ke2017lightgbm} and used the same input features as those described in Section~\ref{sec:input_features}.
Moreover, as ablated variants of \proposedmodel, we considered \proposedmodel with or without Intra- and InterMRA.

\begin{table}[t]
\centering
\caption{Performance comparison on Production Ad-LP dataset. 
}
\begin{tabular}{@{}lrrr@{}}
\toprule
Model                          & \multicolumn{1}{l}{RMSE $\downarrow$} & \multicolumn{1}{l}{MAPE $\downarrow$} & \multicolumn{1}{l}{AUC $\uparrow$} \\ \cmidrule(r){1-1} \cmidrule(lr){2-2} \cmidrule(lr){3-3} \cmidrule(l){4-4}
LightGBM~\cite{ke2017lightgbm} & 1.0000                                & 1.0000                                & 1.0000                             \\ \cmidrule(r){1-1} \cmidrule(lr){2-2} \cmidrule(lr){3-3} \cmidrule(l){4-4}
BERT encoder only              & 0.8615                                & 0.6544                                & 1.2966                             \\
+ IntraMRA                        & 0.8574                                & \textbf{0.6085}                       & 1.3112                             \\
+ InterMRA                        & 0.8591                                & 0.6098                                & 1.3057                             \\
+ Intra \& InterMRA          & \textbf{0.8561}                       & 0.6254                                & \textbf{1.3360}                    \\ \bottomrule
\end{tabular}%
\label{tab:ad_lp_dataset}
\end{table}

Table~\ref{tab:ad_lp_dataset} presents the results of the models in terms of the RMSE, MAPE, and AUC.
To avoid exposing the raw measurements of the production model, the performance of each model was divided by that of the LightGBM model, as indicated at the top of the table.
We considered the RMSE as the main metric for the production scenario, and we were interested in the difference between the true and predicted probabilities because the prediction of the exact probability (i.e., conversion rate) is crucial for a cost-effective ad.
\proposedmodel with Intra- and InterMRA exhibited a substantial performance gain in terms of the RMSE compared to the LightGBM model.
All measures revealed the performance gains of the Inter- and IntraMRA.
We observed the same trend for this dataset as those for the MM-IMDB and Ads-Parallelity datasets, whereas the gain of IntraMRA was larger than that of InterMRA, and \proposedmodel achieved the best performance with both of them.

\subsubsection{Ablation Study on Similarity Function}\label{sec:ablation_similarity_function}

\begin{table}[t]
\centering
\caption{Performance comparison of similarity function variants in MRA.}
\label{tab:additive_vs_scaled_dot_product}
\begin{tabular}{@{}lcrcr@{}}
\toprule
\multicolumn{1}{l}{\multirow{2}{*}{Similarity function}} & \multicolumn{2}{c}{MM-IMDB}                              & \multicolumn{2}{c}{Ads-Parallelity}                 \\ \cmidrule(r){2-3} \cmidrule(l){4-5}
\multicolumn{1}{c}{}                                     & F1-micro                  & \multicolumn{1}{c}{F1-macro} & Acc.                      & \multicolumn{1}{c}{AUC} \\ \cmidrule(r){1-1} \cmidrule(lr){2-2} \cmidrule(lr){3-3} \cmidrule(lr){4-4} \cmidrule(l){5-5}
Additive~\cite{bahdanau2015neural}             & \multicolumn{1}{r}{69.64} & 63.89                        & \multicolumn{1}{r}{78.15} & 79.22                   \\
Scaled dot-product~\cite{vaswani2017attention} & \multicolumn{1}{r}{68.44} & 62.55                        & \multicolumn{1}{r}{78.04} & 77.77                   \\ \bottomrule
\end{tabular}%
\end{table}

Table~\ref{tab:additive_vs_scaled_dot_product} presents a comparison of the \proposedmodel prediction performance with different variations of the similarity function.
We observed an average performance difference of approximately 1\% between the additive attention and scaled dot-product attention on both the MM-IMDB and Ads-Parallelity datasets.
This result confirms that the additive attention in our \proposedmodel is effective.
An extension that replaces MRAs can be considered by adding a Transformer module on top of the BERT encoder.
However, the module is based on scaled dot-product attention, the results suggest that the replacement would exhibit limited performance gains.

\subsubsection{Ablation Study on Residual Connection for MRA}

\begin{table}[t]
    \centering
     \caption{Performance comparison between with and without residual connection in MRA.}
    \begin{tabular}{@{}lcrcr@{}}
        \toprule
        \multirow{2}{*}{\begin{tabular}[c]{@{}l@{}}Residual\\ connection\end{tabular}}  &   \multicolumn{2}{c}{MM-IMDB} &   \multicolumn{2}{c}{Ads-Parallelity} \\
        \cmidrule(r){2-3} \cmidrule(l){4-5}
        &   {\small F1-micro}   &   \multicolumn{1}{c}{\small F1-macro} &   {\small Acc.}   &   \multicolumn{1}{c}{\small AUC}  \\
        \cmidrule(r){1-1} \cmidrule(lr){2-2} \cmidrule(lr){3-3} \cmidrule(lr){4-4} \cmidrule(l){5-5}
        without &   \multicolumn{1}{r}{68.32}   &   62.57   &   \multicolumn{1}{r}{76.70}   &   77.80   \\
        with    &   \multicolumn{1}{r}{\textbf{69.64}}  &   \textbf{63.89}  &   \multicolumn{1}{r}{\textbf{78.15}}  &   \textbf{79.22}  \\
        \bottomrule
    \end{tabular}
   
    \label{tab:residual_connection}
\end{table}

Table~\ref{tab:residual_connection} displays a performance comparison of \proposedmodel with and without a residual connection in the MRA.
In both the MM-IMDB and Ads-Parallelity datasets, the residual connection contributed to an average performance improvement of approximately 2\%.
Therefore, the residual connection plays an important role in MRA.

\subsubsection{Erasing Modality Feature During Inference}

\begin{figure}[t]
    \centering
    \includegraphics[width=\linewidth]{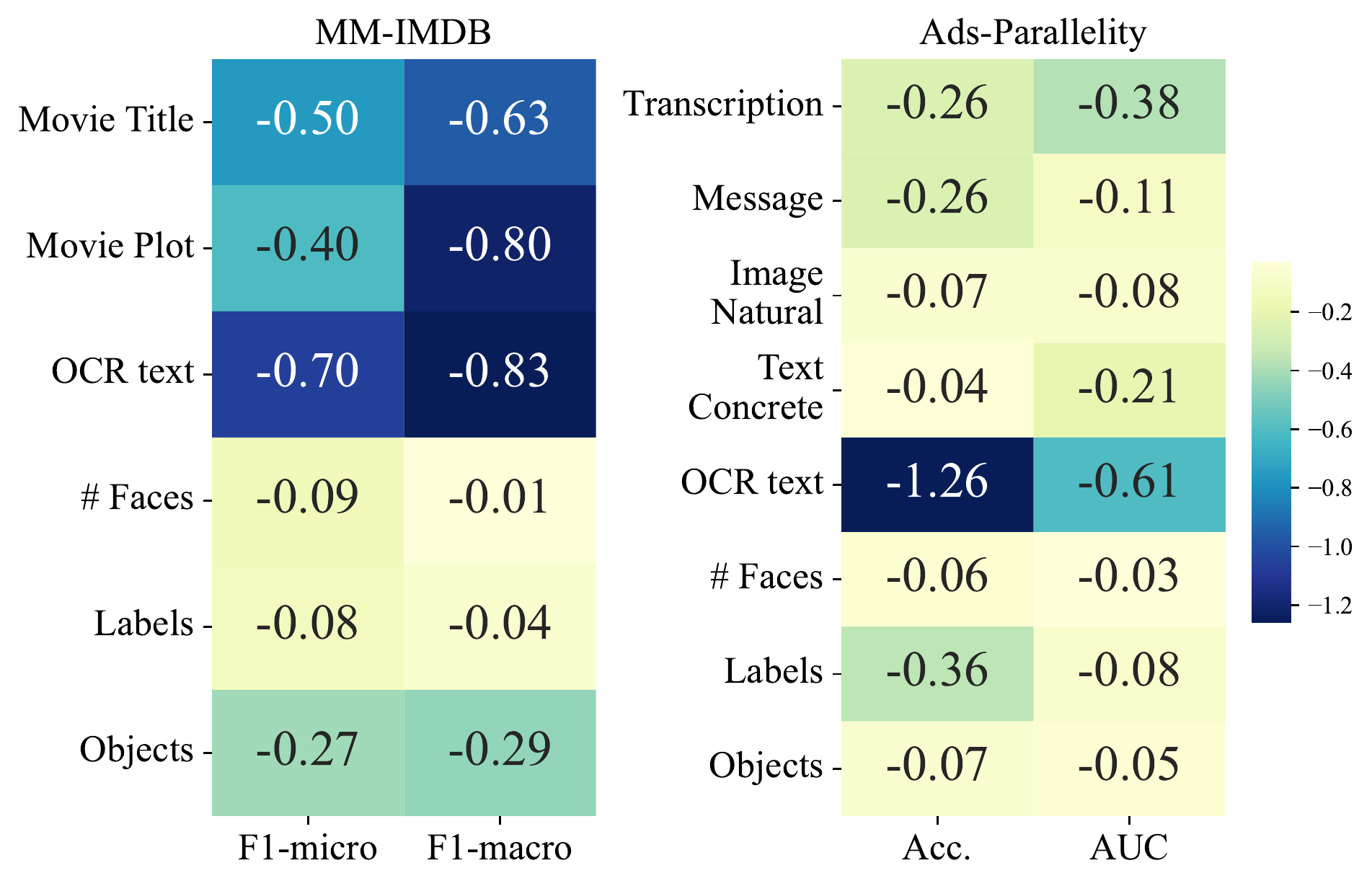}
    \caption{Visualization of the change in performance when each modality was erased during inference. The performance was degraded regardless of which modality feature is erased, and this was particularly significant when erasing OCR text.}
    \label{fig:erasing_features}
\end{figure}

We compared the prediction performance when one of the modalities was erased to identify the modalities that contribute to the prediction in \proposedmodel.
Fig.~\ref{fig:erasing_features} depicts a visualization of the prediction performance of our model when one modality was erased during inference.
We confirmed that erasing any modality feature impaired the performance.
Remarkably, a significant performance degradation was observed when the OCR text was erased.
This result implies that the OCR text that is obtained from the visual modality contributes to the model's prediction.

\subsubsection{Visualization of Inter and IntraMRA}

\begin{figure*}[t]
    \centering
    \includegraphics[width=0.97\linewidth]{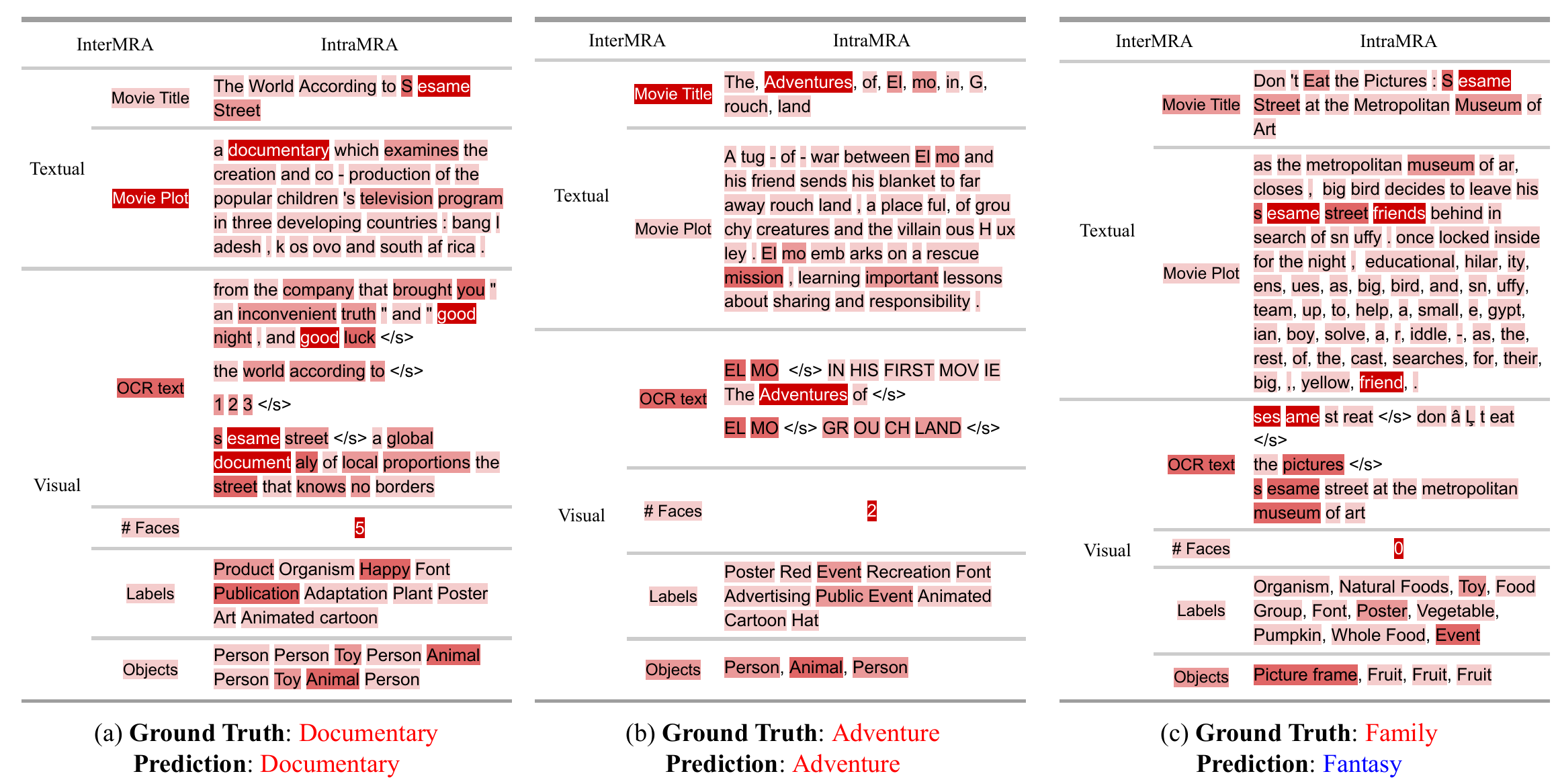}
    \caption{Visualization results for the attention weights of Inter- and IntraMRA in MM-IMDB dataset. A deeper red indicates activation.}
    \label{fig:visualization_result}
\end{figure*}

Fig.~\ref{fig:visualization_result} presents the visualization results for the attention weights of the Inter- and IntraMRA on the MM-IMDB dataset.
In Fig.~\ref{fig:visualization_result} (a) and \ref{fig:visualization_result} (b), where the predictions were correct, the MRA captured the corresponding words (or subwords) that contributed to the prediction.
For example, in Fig.~\ref{fig:visualization_result} (a), we can observe a strong response to words relating to the documentary film in IntraMRA (e.g., ``documentary'' in movie plot).
Furthermore, InterMRA could select important modalities; for example, it strongly focused on movie plot and OCR text features that contained words directly relating to prediction.
Remarkably, it can be observed from Fig.~\ref{fig:visualization_result} (a) that high IntraMRA weights were assigned to the split words ``document'' and ``ary'' in the OCR text.
This is possibly because the BERT-based encoder encoded the two words (i.e., ``document'' and ``ary'') by considering the context, and IntraMRA could assign importance to them based on the encoded representations.
Consequently, the errors in the OCR could be compensated.
This can be viewed as evidence for that, even with EasyOCR, \proposedmodel can achieve comparable performance to that of the DMMS with a proprietary OCR method (see Tables~\ref{tab:sota_mm_imdb} and \ref{tab:sota_ads_parallelity}).
In the samples in which the model made an incorrect prediction (as illustrated in Fig.~\ref{fig:visualization_result} (c)), 
although the model could capture important words such as ``friends'' in the movie plot, the overall importance was dominated by the other modalities.

\section{Conclusion and Future Work}
    We have proposed a new concept for a multimodal set-input DNN architecture, namely \proposedmodel.
Our proposed concept consists of three projections that capture the global and local relationships among multiple modalities.
\proposedmodel~is specifically designed for sequence sets, which are the structured data that are often collected in real multimodal applications.
Whereas \proposedmodel~is compatible with the sequential embeddings that are provided by recent BERT-based encoders, the Intra- and InterMRA capture the hierarchical importance of the elements in the modality sequences.
We empirically demonstrated the effectiveness of our concept compared to state-of-the-art multimodal models.
Furthermore, we examined its applicability to real applications by confirming the substantial performance gain from production-running models using real-world Ad data.

In future work, we would like to explore attention mechanisms for various structures of multimodal data, such as grouped modalities (i.e., sets of sets) and hierarchical fields (i.e., sets of trees), further.
Moreover, we are planning to investigate the theoretical relationship between our attention-based fusion and conventional multimodal methods, such as gradient blending~\cite{wang2020makes} and factorization machines~\cite{rendle2010factorization}.
    
\appendices
\begin{table*}[t]
    \centering
    \caption{Statistics of the Production Ad-LP dataset.}
    \begin{threeparttable}
    \begin{tabular}{@{}clllr@{}}
        \toprule
        \multicolumn{3}{c}{Modality}    & \multicolumn{1}{c}{\multirow{2}{*}{Modality Description}} &  \multicolumn{1}{c}{\multirow{2}{*}{Details}}    \\
        \cmidrule(r){1-3}
        \multicolumn{1}{l}{High-level modality} &   \multicolumn{2}{c}{Low-level modality}  &   \multicolumn{1}{c}{}    & \multicolumn{1}{c}{}  \\
        \cmidrule(r){1-1} \cmidrule(lr){2-3} \cmidrule(lr){4-4} \cmidrule(l){5-5}
        \multirow{9}{*}{Ad} &   Keyword &   &   Search Keywords &   2.22$\pm$1.07 words \\
        \cmidrule(l){2-3} \cmidrule(lr){4-4} \cmidrule(l){5-5}
        & \multirow{3}{*}{Title}    &   $\texttt{title\_1}$ &   \multirow{3}{*}{Title text of the Ad}   &  6.43$\pm$2.02 words \\
        &   &   $\texttt{title\_2}^{\dagger}$   &   &   6.90$\pm$2.25 words \\
        &   &   $\texttt{title\_3}^{\dagger}$   &   &   3.95$\pm$3.51 words \\
        \cmidrule(l){2-3} \cmidrule(lr){4-4} \cmidrule(l){5-5}  &   \multirow{2}{*}{Description}    &   $\texttt{description\_1}$  &   \multirow{2}{*}{Description text of the Ad} & 20.02$\pm$4.28 words   \\
        &   &   $\texttt{description\_2}^{\dagger}$ &   &   12.23$\pm$10.05 words   \\
        \cmidrule(l){2-3} \cmidrule(lr){4-4} \cmidrule(l){5-5} 
        &   \multirow{2}{*}{Path}   &   $\texttt{path\_1}^{\dagger}$    &   \multirow{2}{*}{URL path text}   &   1.03$\pm$1.14 words \\
        &   &   $\texttt{path\_2}^{\dagger}$    &   &   2.23$\pm$1.23 words \\
        \cmidrule(l){2-3} \cmidrule(lr){4-4} \cmidrule(l){5-5} 
        &   Match type  &   &   Match type for the search keywords  &   3 types \\
        \midrule
        \multirow{6}{*}{LP} &   LP text &   &   \begin{tabular}[c]{@{}l@{}}Bounding box feature obtained by OCR \\ from a screenshot of the LP\end{tabular}  &   223.87$\pm$102.83 words \\
        \cmidrule(l){2-3} \cmidrule(lr){4-4} \cmidrule(l){5-5} 
        &   Safe search &   &   \begin{tabular}[c]{@{}l@{}}Explicit content such as adult content \\ or violent content within an image.$^{1}$\end{tabular} &   5 types \\
        \cmidrule(l){2-3} \cmidrule(lr){4-4} \cmidrule(l){5-5} 
        &   Label detection &   &   \begin{tabular}[c]{@{}l@{}}Information about entities in an image, \\ across a broad group of categories.$^{2}$\end{tabular}   &   \multirow{2}{*}{1,782 categories} \\
        \cmidrule(lr){2-3} \cmidrule(lr){4-4}
        &   Object detection    &   &   \begin{tabular}[c]{@{}l@{}}Multiple objects in an image \\ with Object Localization.$^{3}$\end{tabular} &   \\
        \cmidrule(l){2-3} \cmidrule(lr){4-4} \cmidrule(l){5-5} 
        &   \begin{tabular}[c]{@{}l@{}}Number of faces \\ in the LP\end{tabular}    &   &   \begin{tabular}[c]{@{}l@{}}Number of faces in the screenshot \\ based on Face Detection API$^{4}$\end{tabular}  &   1.13$\pm$2.22   \\
        \cmidrule(l){2-3} \cmidrule(lr){4-4} \cmidrule(l){5-5} 
        &   \begin{tabular}[c]{@{}l@{}}Lighthouse \\ performance\end{tabular}   &   &   Lighthouse performance of the LP$^{5}$ &   35.22$\pm$20.10 \\
        \midrule
        Ad group ID  &   &   &   &   15,244  \\
        \bottomrule
    \end{tabular}%
    \begin{tablenotes}
    \item[$\dagger$] Optional value ~~~ $^1$ \url{https://cloud.google.com/vision/docs/detecting-safe-search}
    \item[2] \url{https://cloud.google.com/vision/docs/labels} ~~~ $^3$ \url{https://cloud.google.com/vision/docs/object-localizer}
    \item[4] \url{https://cloud.google.com/vision/docs/detecting-faces} ~~~ $^5$ \url{https://developers.google.com/web/tools/lighthouse}
    \end{tablenotes}
    \end{threeparttable}
    \label{tab:dataset}
\end{table*}

\section{The Details of the Production Ad-LP Dataset}\label{sec:details_of_production_ad_lp}

The statistics for the Production Ad-LP dataset are listed in Table~\ref{tab:ad_lp_dataset}.
The dataset was collected from CyberAgent Inc.\footnote{\url{https://www.cyberagent.co.jp/en/corporate/}} from August 1, 200, to November 30, 2020.
This Japanese dataset contains 257,235 search engine advertisements (Ad) and landing pages (LPs), and these two modalities are called high-level modalities.
Each high-level modality includes some low-level modalities.
For example, the Ad modality contains search keywords, multiple titles (\texttt{title\_1}, \texttt{title\_2}, and \texttt{title\_3}, where \texttt{title\_2} and \texttt{title\_3} are optional values), multiple descriptions of the ads (\texttt{description\_1} and \texttt{description\_2}, where \texttt{description\_2} is an optional value), an LP URL path (\texttt{path\_1} and \texttt{path\_2}, where both are optional), and a match type.\footnote{About keyword matching options - Google Ads Help \url{https://support.google.com/google-ads/answer/7478529}}
Several low-level modalities exist in the LP, such as the LP text that is obtained by applying optical character recognition (OCR) from LP screenshot images and safe search results (five types: adult, spoof, medical, violence, and racy), labels, object face detection results, and page performance scores with the lighthouse tool.\footnote{\url{https://developers.google.com/web/tools/lighthouse}}
The statistics of the features of each row are shown in the Details column.
We used a screenshot of the first view of the landing page at iPhone 8 resolution (1334 $\times$ 750).
The lighthouse performance has a value between 0 and 100.

We divided the dataset in an Ad group-based manner to train and evaluate the models.
The Ad group\footnote{Ad group: Definition - Google Ads Help \url{https://support.google.com/google-ads/answer/6298?hl=en}} contained one or more ads that shared similar targets.
In most advertising systems, ads are served in campaign units, and multiple creatives with similar trends are developed during a campaign.
Each ad campaign consists of one or more ad groups.
Thus, to avoid data leakage owing to potential similarity, we divided the dataset into a training set, a validation set, and test sets in an ad group-based manner.

Note that, as our experiments were based on open datasets that are widely available to the public for performance comparison and results from analysis, no privacy handling issues would occur.
In addition to the advertising data received from the company, information that could identify individuals was also discarded.
Moreover, we took great care to ensure that the results of our analysis would not be disclosed in any manner that would violate privacy or ethics.

\section{Implementation Details}\label{sec:appendix_implementation_details}

We used the OCR\footnote{\url{https://cloud.google.com/vision/docs/ocr}}, label detection\footnote{\url{https://cloud.google.com/vision/docs/labels}}, object detection\footnote{\url{https://cloud.google.com/vision/docs/object-localizer}}, and facial detection\footnote{\url{https://cloud.google.com/vision/docs/detecting-faces}} services to obtain visual features as tokens in the Google Cloud Vision API.
We converted the obtained features, including numeric ones, as strings and treated them as tokens.

We used the pre-trained RoBERTa~\cite{liu2019roberta} large encoder\footnote{\url{https://huggingface.co/roberta-large}} in MM-IMDB and Ads-Parallelity following Reiter~\textit{et al.}~\cite{reiter2020deep} as the BERT-based encoder.
This encoder outputs a hidden state with dimension $d = 1024$.
In the Production Ad-LP dataset, we used a pre-trained Japanese BERT\footnote{\url{https://huggingface.co/cl-tohoku/bert-base-japanese-whole-word-masking}} as the encoder.
This encoder outputs a hidden state with dimension $d = 768$.

To segment the Japanese text of the ads into words in the Ad-LP dataset, we used MeCab~\cite{kudo2006mecab}, which is a type of morphological analyzer.
We used the analysis dictionary for the morphological analyzer MeCab (UniDicMA), as the custom dictionary.
These are in accordance with the standard methods of the Japanese BERT.

\section*{Acknowledgment}
We would like to appreciate the editors and anonymous reviewers for their helpful feedback.
We also thank the CyberAgent AI Lab team, especially Kota Yamaguchi and Daisuke Moriwaki, for their insightful comments.

\bibliographystyle{IEEEtran}
\bibliography{references}

\EOD

\end{document}